\documentclass[]{article}

\usepackage{multirow, amsmath, amsthm, graphicx, amssymb, amsfonts, latexsym, enumerate, amscd, xcolor, float, caption, bbold}
\usepackage{listings}
\usepackage{geometry}
\RequirePackage[numbers]{natbib}

\DeclareMathOperator{\sign}{sign}
\DeclareMathOperator{\diag}{diag} 
\DeclareMathOperator{\G}{G} 
\DeclareMathOperator{\E}{E} 
\DeclareMathOperator{\Var}{Var} 
\DeclareMathOperator{\SN}{SN} 
\DeclareMathOperator{\ESN}{ESN}
\DeclareMathOperator{\erf}{erf}

\usepackage[title]{appendix}

\usepackage{etoolbox}
\apptocmd{\appendices}{\apptocmd{\thesection}{:}{}{}}{}{}

\begin{document}

\title{An Extended Simplified Laplace strategy for Approximate Bayesian inference of Latent Gaussian Models using R-INLA}

\author{Cristian Chiuchiolo (cristian.chiuchiolo@kaust.edu.sa )\\
AND \\
Janet van Niekerk (janet.vanniekerk@kaust.edu.sa ) \\
AND \\
H\aa vard Rue (haavard.rue@kaust.edu.sa )\\
CEMSE Division\\
	King Abdullah University of Science and Technology\\
	Kingdom of Saudi Arabia}

\maketitle

\begin{abstract}

Various computational challenges arise when applying Bayesian inference approaches to complex hierarchical models. Sampling-based inference methods, such as Markov Chain Monte Carlo strategies, are renowned for providing accurate results but with high computational costs and slow or questionable convergence. On the contrary, approximate methods like the Integrated Nested Laplace Approximation (INLA) construct a deterministic approximation to the univariate posteriors through nested Laplace Approximations. This method enables fast inference performance in Latent Gaussian Models, which encode a large class of hierarchical models. R-INLA software mainly consists of three strategies to compute all the required posterior approximations depending on the accuracy requirements. The Simplified Laplace approximation (SLA) is the most attractive because of its speed performance since it is based on a Taylor expansion up to order three of a full Laplace Approximation. Here we enhance the methodology by simplifying the computations necessary for the skewness and modal configuration. Then we propose an expansion up to order four and use the Extended Skew Normal distribution as a new parametric fit. The resulting approximations to the marginal posterior densities are more accurate than those calculated with the SLA, with essentially no additional cost.

\end{abstract}

\section{Introduction}

Hierarchical models appear to be challenging within a Bayesian inference framework due to their highly correlated structure and/or high dimensionality. Sampling-based methods such as Markov Chain Monte Carlo (MCMC) may require non-negligible computational demand when applied to these models. Approximate methods such as Laplace approximations aim to circumvent the high computational demand of sampling-based methods by approximating marginal posterior distributions \cite{ventura2014, ventura2016, ventura2016l}. A simple Gaussian approximation (Laplace Method) for the unknown joint density parameters can be quite crude and restricts the marginal posterior densities to be symmetric. Integrated Nested Laplace Approximations (INLA) as introduced by \cite{rue2009} are based on a series of Laplace approximations for the regression parameters, for example, resulting in an unknown parametric form of the marginal posterior density functions. Nonetheless, INLA performs full Bayesian inference in a fraction of the time of sampling-based methods. \\ \\
A Simplified Laplace strategy (SLA) was proposed as a simpler alternative to the original INLA, in the sense that Skew Normal densities approximate the marginal posterior density functions \cite{Azzalini_1999,Azzalini_2003, canale2011, canale2015, seijas2017presence, Perez2017, simon2020}. This approach implies a skew marginal posterior density, which improves the Gaussian marginals from the Laplace method. The Skew Normal approximations capture the true marginal posterior densities quite accurately, while the computational cost for this approach is much less than the full INLA. This Skew Normal family-based approach embodies the optimal strategy to approximate skewed marginal and joint posterior densities \cite{egil2015}. \\ \\
We propose a new approximation for the marginals of unknown latent parameters of a latent Gaussian model based on the INLA framework. We use an extended Skew Normal approximation to the marginal posterior densities using the Extended Skew Normal distribution \cite{adelchiazzalini_2018_the}, with the hope of capturing the skewness and kurtosis more accurately, than the Skew Normal approximations. We term this approximation an Extended Simplified Laplace Approximation (ESLA), as a direct extension of the Simplified Laplace strategy employed by the \textit{R-INLA} R package. This extension is a first attempt to move beyond the Skew Normal density approximation, while maintaining computational efficiency. \\ \\
In Section \ref{sec:1} we discuss the Latent Gaussian Model formulation by emphasizing the role of Gaussian assumptions onto the latent field, which appear to be a natural choice when a deterministic approach such as INLA is used. Section \ref{sec:2} introduces one possible extension of the Skew Normal distribution, the Extended Skew Normal, as a candidate for approximating the marginals when marginal skewness is non-negligible. Section \ref{sec:3} presents a transparent and wieldy way to localize the mode when fitting Skew Normal distributions, whereafter we present the details of the extended Skew Normal approximation. Section \ref{sec:4} shows some general skewed examples where the new strategy is applied and compared to the other approaches. Section \ref{sec:5} contains a brief discussion on the proposed methodology and its performance as well as possibilities for further extensions in this regard.

\section{Latent Gaussian models (LGM) and Integrated Nested Laplace Approximations (INLA)}
\label{sec:1}

\subsection{LGM}

Latent Gaussian Models (LGMs) are appealing in Bayesian computational inference when using INLA for two main intrinsic assumptions: the log-likelihood contribution is log concave in terms of its linear predictor and the latent field is Gaussian distributed, a priori. These assumptions ensure the posterior distribution of the model to be Gaussian-like and therefore easily handled by INLA. Log concavity on the log-likelihood is a strong beneficial assumption as it enforces the distribution on the observed data to be close to a Gaussian distribution when conditional independence to each latent term is assumed. \\ Such assumptions make clear that it is less fruitful to assume a statistical structure that goes far beyond a Gaussian distribution when dealing with Latent Gaussian Models. As an example, we consider a simple latent structure with no hyperparameters. By model assumptions we have the latent field $\boldsymbol{x} \sim \text{N}(\boldsymbol{0}, \boldsymbol{Q})$ with a precision matrix $\boldsymbol{Q}$ having marginal variances equal to 1, and likelihood contribution $\boldsymbol{y} \vert \boldsymbol{x} \sim \prod_i \pi(y_i \vert x_i)$ with $n$ data observations. \\ We assume $\vert \pi(y_i \vert x_i) \vert < \tilde{C}_i $ where each likelihood density is a function of $x_i$ bounded by a constant $\tilde{C}_i$ which is unique for each observation. The posterior distribution of the corresponding latent model is  

\begin{equation}
\pi(\boldsymbol{x} \vert \boldsymbol{y}) \propto \pi(\boldsymbol{x}) \pi(\boldsymbol{y} \vert \boldsymbol{x}) \le \pi(\boldsymbol{x}) \tilde{C}
\label{post_const_gau}
\end{equation}
where $\tilde{C}= \prod_i \tilde{C}_i$. Since this Gaussian bound exists for the latent density, we can question if a similar bound is preserved for each latent marginal. We will show that

\begin{equation}
\pi(\boldsymbol{x} \vert \boldsymbol{y}) \le \tilde{C} \pi(\boldsymbol{x}) \Rightarrow \pi(x_i \vert \boldsymbol{y}) \le \tilde{C} \pi(x_i)
\label{gau_arg}
\end{equation}
where $\pi(x_i)$ is the respective $i^{th}$ Gaussian marginal density from its multivariate counterpart $\pi(\boldsymbol{x})$. \\ The above statement provides a legitimate justification to using Gaussian assumptions onto the latent field of a Latent Gaussian Model structure. This marginal implication can be shown in few steps. \\ We define functions $g_i(x_i)= \log(\pi(y_i \vert x_i))$ and write the latent joint conditional density as

\begin{equation}
\pi(\boldsymbol{x} \vert \boldsymbol{y}) = G \exp \Bigl ( -\frac{1}{2} \boldsymbol{x}^T \boldsymbol{Qx}+\sum_{i=1}^n g_i(x_i) \Bigr )
\end{equation}

\noindent
where $G$ is the normalization constant. Each posterior latent marginal $x_i$ is obtained by integrating out all the other latent components $\boldsymbol{x}_{-i}$ 

\begin{align}
\pi(x_i \vert \boldsymbol{y}) &= \int_{\boldsymbol{x}_{-i}} \pi(\boldsymbol{x} \vert \boldsymbol{y}) \, d \boldsymbol{x}_{-i} \nonumber \\
&= \exp (g_i(x_i)) \int_{\boldsymbol{x}_{-i}} G\exp \Bigl ( -\frac{1}{2} \boldsymbol{x}^T \boldsymbol{Qx} \Bigr ) \exp \Bigl ( \sum_{j \ne i} g_j(x_j) \Bigr ) \, d \boldsymbol{x}_{-i}
\end{align}
Since each $g_i(x_i)$ is bounded per our initial assumptions, then 

\begin{align}
\pi(x_i \vert \boldsymbol{y}) &\le \tilde{C}_i \int_{\boldsymbol{x}_{-i}} G\exp \Bigl ( -\frac{1}{2} \boldsymbol{x}^T \boldsymbol{Qx} \Bigr ) \prod_{j \ne i} \tilde{C}_j \, d \boldsymbol{x}_{-i} \nonumber \\
& \le \tilde{C} \exp \Bigl ( -\frac{1}{2} x_i^2(\boldsymbol{Q}^{-1})_{ii}^{-1} \Bigr )
\label{gau_final}
\end{align}
which corresponds to~\eqref{gau_arg}. The notation $(\boldsymbol{Q}^{-1})_{ii}$ refers to the $i^{th}$ marginal variance term $\boldsymbol{\Sigma}_{ii}$ derived from the covariance matrix $\boldsymbol{\Sigma}=\boldsymbol{Q}^{-1}$. The result~\eqref{gau_final} shows that the Gaussian distribution represents a natural bound for each marginal up to a constant. This emphasizes that distributions with a Gaussian-like behavior are the most natural choice for approximating posterior marginals. Their tails must follow a Gaussian behavior while the main bulk of the distribution can show differences from a Gaussian density because of location and skewness. \\ The Gaussian and Simplified Laplace strategies represent an appropriate embodiment of this Gaussian feature since their application provides accurate marginal posterior approximations in most of the cases by exploiting Gaussian-like distributions. In Section \ref{sec:2} we show that Skew Normal family distributions are natural candidates as their tail behavior approximately resembles the one from a Gaussian distribution. \\ Appendix A also discusses if this argument still holds when considering more heavy-tailed assumptions such as the Student-t distribution. 

\subsection{INLA}

The Integrated Nested Laplace Approximation (INLA) methodology consists of computing discrete approximations to univariate posteriors of the unknown parameters of a Latent Gaussian Model (LGM). Amongst others, the Stochastic Partial Differential Equation (SPDE) approach employed by the INLA methodology in the geostatistics field \cite{lindgren2011} has heavily impacted the applied sciences community. New insights and extensions about the interpolation algorithms applied to the hyperparameter posterior marginals to improve speed while retaining accuracy is presented by \cite{tmartins2013, simpson2011fast}. Enhanced model features \cite{tmartins2014} and GLMMs corrections \cite{egil2015}, a measurement error model \cite{muff2015}, introduction of a new prior methodology \cite{simpson2017penalising}, criticisms and Bayesian model diagnostics \cite{egil2017}, a book about spatial and spatiotemporal models \cite{blangiardo2013} with more advanced examples in \cite{elias2018}, are all contributions to the INLA methodology and applications. Three main reviews about new advancements can be read at \cite{rue2017, bakka2018, martino2019integrated}. Recent new applications on joint models using the PARDISO library  (\cite{SCHENK2004475}) was proposed by \cite{van2019new, niekerk2021competing}. \\ Clearly, the INLA methodology provides a new path for Bayeisan inference that is efficient, accurate and can be applied to many statistical applications. Here we briefly explain the INLA methodology.\\ \\ Assuming $n$-dimensional data $\boldsymbol{y}$ with likelihood $\prod_i \pi(y_i \vert x_i,\boldsymbol{\theta})$, an unobserved latent field vector $\boldsymbol{x}$ with prior $\pi(\boldsymbol{x} \vert \boldsymbol{\theta})$, and a hyperparameter set $\boldsymbol{\theta}$ with prior $\pi(\boldsymbol{\theta})$, the unknown parameters (latent and hyperparameters) of a Latent Gaussian Model has joint posterior density

\begin{equation}
    \pi(\boldsymbol{x},\boldsymbol{\theta} \vert \boldsymbol{y}) \propto \pi(\boldsymbol{\theta}) \pi(\boldsymbol{x} \vert \boldsymbol{\theta}) \prod_i \pi(y_i \vert x_i,\boldsymbol{\theta}) 
    \label{lgm_post}
\end{equation}
whose implicit hierarchical structure is summarised into

\begin{align}
    \boldsymbol{y} \vert \boldsymbol{x}, \boldsymbol{\theta} &\sim \prod_{i=1}^n \pi(y_i \vert x_i,\boldsymbol{\theta}) \nonumber \\
    \boldsymbol{x} \vert \boldsymbol{\theta} &\sim N(\boldsymbol{0}, \boldsymbol{Q}^{-1}(\boldsymbol{\theta})) \nonumber \\
    \boldsymbol{\theta} &\sim \pi(\boldsymbol{\theta})
\end{align}
The likelihood contribution to the model entirely comes from each $\pi(y_i \vert x_i,\boldsymbol{\theta})$ where each observation $y_i$ only correspond to one single latent term $x_i$. Each observation $y_i$ has a corresponding linear predictor term $\eta_i$ additive for all unknown model parameters: fixed coefficients or random terms related to cluster effects, non-linear functions, temporal or spatial specification. \\ Both these parameters and the linear predictor vector $\boldsymbol{\eta}$ belong to the latent field $\boldsymbol{x}$ which is assumed to be Gaussian distributed with a sparse precision matrix $\boldsymbol{Q}$. This latent Gaussian assumption is well specified for both small and large dimensions by the concept of Gaussian Markov Random Fields (GMRFs, \cite{held2005}) and is fundamental for Latent Gaussian Models. \\ GMRFs allow modeling the dependency structure of the latent components of the model simultaneously, providing the ground for fast computations due to the precision sparsity structure. By encoding the linear predictor into the latent field, INLA can compute all the possible posteriors of the model without much computational effort in most cases. \\ The hyperparameter set $\boldsymbol{\theta}$ contains all the hyperparameters of the Latent Gaussian Model, and its dimension can lead to more costly computations if the dimension of $\boldsymbol{\theta}$ is too high. We can deal with most of the cases routinely when $\vert \boldsymbol{\theta} \vert < 20$. The hyperpriors $\pi(\boldsymbol{\theta})$ are not bounded to be Gaussian, and many different distributions can be used. \\ \\  INLA is a deterministic algorithm that computes accurate approximations for the univariate posterior marginals of the unknown parameters of a Latent Gaussian Model. \\  From the joint posterior density in~\eqref{lgm_post} we derive its marginal densities as follows

\begin{align}
    \pi(x_i \vert \boldsymbol{y})=\int_{\boldsymbol{\theta}} \pi(x_i \vert \boldsymbol{\theta}, \boldsymbol{y}) \pi(\boldsymbol{\theta} \vert \boldsymbol{y}) \, d \boldsymbol{\theta}, \quad i=1,\dots,N 
    \label{inla_mixt} \\
    \pi(\theta_j \vert \boldsymbol{y})= \int_{\boldsymbol{\theta}_{-j}} \pi(\boldsymbol{\theta} \vert \boldsymbol{y}) \, d \boldsymbol{\theta}_{-j}, \quad j=1,\dots,p
    \label{inla_marg}
\end{align}
with $N$ being the overall dimension of the latent field $\boldsymbol{x}$ and $p$ being the dimension of the hyperparameter $\boldsymbol{\theta}$. The approximations of the marginals in~\eqref{inla_marg} result from numerically integrating out the hyperparameter uncertainty $\boldsymbol{\theta}$ to get

\begin{align}
    \tilde{\pi}(x_i \vert \boldsymbol{y}) \approx \sum_{k=1}^K \tilde{\pi}(x_i \vert \boldsymbol{y}, \boldsymbol{\theta}_k) \tilde{\pi}(\boldsymbol{\theta}_k \vert \boldsymbol{y}) \Delta_k \nonumber \\
    \tilde{\pi}(\boldsymbol{\theta} \vert \boldsymbol{y}) \propto \frac{\pi(\boldsymbol{x}^*, \boldsymbol{\theta} \vert \boldsymbol{y})}{\tilde{\pi}_{\G}(\boldsymbol{x}^* \vert \boldsymbol{\theta},\boldsymbol{y})} \Bigg \vert_{\boldsymbol{x}^*=\boldsymbol{\mu(\theta)}}
    \label{INLA_approx}
\end{align}
with $K$ being the total number of points used in the numerical integration process (see \cite{rue2009, martino2019integrated}). $\tilde{\pi}_{\G}(\boldsymbol{x}^* \vert \boldsymbol{\theta},\boldsymbol{y})$ is the Gaussian approximation obtained by matching the mode and curvature at the mode of the full joint density $\pi(\boldsymbol{x} \vert \boldsymbol{\theta}, \boldsymbol{y})$ found after an iterative process. \\ The whole methodology can be summarised as follows

\begin{description}
   \item[$\bullet$] Explore the approximation $\log \tilde{\pi}(\boldsymbol{\theta} \vert \boldsymbol{y})$ through a grid exploration scheme in a $p$-dimensional space. Locate the mode and compute a set of configuration points $\boldsymbol{\theta}_k, k=1\dots,K$ within the region oh high probability mass
   \item[$\bullet$] Evaluate $\log \tilde{\pi}(\theta_1 \vert \boldsymbol{y}), \dots, \log \tilde{\pi}(\theta_K \vert \boldsymbol{y})$ and use these results to compute both $\tilde{\pi}(x_i \vert \boldsymbol{y}, \boldsymbol{\theta}_k)$ and $\tilde{\pi}(x_i \vert \boldsymbol{y}) $
\end{description}
INLA computes first the hyperparameter posterior marginals in~\eqref{INLA_approx} by using Laplace Approximations on the entire ratio and then evaluates it at the denominator mean $\boldsymbol{\mu(\theta)}$. A second Laplace Approximation is then applied to the full conditional posterior densities $\pi(x_i \vert \boldsymbol{y}, \boldsymbol{\theta}_k)$ by using the pre-computed points $\boldsymbol{\theta}_k$ as follows

\begin{equation}
    \tilde{\pi}(x_i \vert \boldsymbol{y}, \boldsymbol{\theta}_k) \approx \frac{\pi(\boldsymbol{x}^*, \boldsymbol{\theta}_k \vert \boldsymbol{y})}{\tilde{\pi}_{\G}(\boldsymbol{x}_{-i}^* \vert x_i, \boldsymbol{\theta}_k,\boldsymbol{y})} \Bigg \vert_{\boldsymbol{x}_{-i}^*=\boldsymbol{\mu}_{-i}(\boldsymbol{\theta)}}
    \label{full_approx}
\end{equation}
with $\tilde{\pi}_{\G}(\boldsymbol{x}_{-i}^* \vert x_i, \boldsymbol{\theta}_k,\boldsymbol{y})$ being the Gaussian Approximation with modal configuration $\boldsymbol{\mu}_{-i}(\boldsymbol{\theta})$. \\ Depending on the nature of these densities, there are three approximation strategies that can be applied to~\eqref{full_approx} (in order of increasing accuracy and computational cost): \emph{Gaussian Approximation}, \emph{Simplified Laplace Approximation} and \emph{Laplace Approximation}. \\ The Gaussian approximation is preferable when the Gaussian assumptions hold for the model likelihood and has the best speed performance. However, this strategy may have inaccuracies in location and skewness adjustments when the likelihood contribution deviates significantly from a Gaussian behavior (as also stated in \cite{martino2007}). \\ On the other side, the Laplace approximation is more computationally intensive but ensures more accuracy since it exploits a more on-point Gaussian approximation for each latent term at the denominator in~\eqref{full_approx}. Then the ratio is simplified through a series of selected points based on selected criteria that save computations (see \cite{rue2009} for details), and the marginal approximated density result for each latent term is given by the product of a Gaussian kernel and a cubic spline. The spline itself interpolates selected points of the marginal latent variable to the log density difference between the resulting Laplace approximation and respective Gaussian approximation. Another strategy available is the Simplified Laplace approach, which applies a third-order Taylor expansion of the Laplace approximation, therefore providing a more computational-friendly result at some negligible accuracy in most cases. This strategy is the INLA default choice and exploits Skew Normal densities to get the full conditional approximations $\pi(x_i \vert \boldsymbol{y}, \boldsymbol{\theta}_k) $ in~\eqref{full_approx}. Most of the present work goes through the details of the Simplified Laplace approximations while proposing a way to extend its capabilities through another distribution of the Skew Normal class. 

\section{The Extended Skew Normal Distribution and its properties}
\label{sec:2}

INLA uses the Skew Normal family when the Gaussian assumptions are not accurate enough. These skewed distributions tend to be good approximations of the marginal posteriors of a Latent Gaussian Model. \\ Observed skewness is retrieved through the third moment of a Skew Normal distribution. In more extreme settings, the marginal skewness from the full conditional densities $\pi(x_i \vert \boldsymbol{\theta}, \boldsymbol{y})$ can benefit from a more complex structure with additional free parameters. \\ The Extended Skew Normal distribution (see \cite{Azzalini_1999} for other alternative distributions and more insights on the Skew Normal family) belongs to the Skew Normal family and can model skewness using two parameters instead of one. \\ First, we introduce some basic definitions and properties of this extended version of the Skew Normal family. We define $T \sim \text{ESN}(\xi, \omega, \alpha, \tau)$ to be an Extended Skew Normal random variable whose probability density function is

\begin{equation}
f(t; \xi, \omega, \alpha, \tau) = \frac{1}{\omega \Phi(\tau)} \phi \Bigl ( \frac{t-\xi}{\omega}\Bigr ) \Phi \Bigl ( \tau \sqrt{\alpha^2+1}+ \alpha \frac{t-\xi}{\omega}\Bigr ) 
\label{esn}
\end{equation}

\noindent
with location parameter $\xi$, scale $\omega$, skewness parameter $\alpha$ and hidden mean parameter $\tau$ (or truncation parameter as mentioned in \cite{canale2011, adelchiazzalini_2018_the}) while $\phi(\cdot)$, $\Phi(\cdot)$ are respectively the probability and cumulative density function of a standard Gaussian. \\ For $\tau=0$ the equation in~\eqref{esn} reduces to a Skew Normal distribution with the same parameter notation.  The cumulant generating function of $T$ is given by

\begin{equation}
K(u)= \log M(u)=\xi u + \frac{1}{2} \omega^2 u^2 + \mathcal{C}_0 (\tau+\delta \omega u)-\mathcal{C}_0(\tau)
\end{equation}

\noindent
where $M(u)=\E[e^{uT}]$ is the moment generating function with parameterization $\delta = \frac{\alpha}{\sqrt{1+\alpha^2}}$ and $\mathcal{C}_0 (z)= \log 2 \Phi(z)$. From $K(u)$ we get the first four moments 

\begin{align}
\E(T) &= \xi+\mathcal{C}_1 (\tau)\omega \delta \nonumber \\
\Var(T) &= \omega^2 [1+\mathcal{C}_2 (\tau) \delta^2 ] \nonumber \\
\gamma_1 (T) &= \frac{\mathcal{C}_3(\tau)\delta^3}{(1+\mathcal{C}_2 (\tau) \delta^2)^{3/2}} \nonumber \\
\gamma_2 (T) &= \frac{\mathcal{C}_4(\tau)\delta^4}{(1+\mathcal{C}_2 (\tau) \delta^2)^2}
\label{esn_mom}
\end{align}

\noindent
with $\gamma_1$, $\gamma_2$ being the standardized skewness and kurtosis. The $\mathcal{C}(\cdot)$ functions are defined with respect to $\tau$ by \cite{adelchiazzalini_2018_the} as

\begin{equation}
\mathcal{C}_r(\tau)= \frac{\partial^r}{\partial \tau^r} \log 2\Phi(\tau)
\end{equation}

\noindent
with the first five derivatives being

\begin{align}
\mathcal{C}_1(\tau) &= \frac{\phi(\tau)}{\Phi(\tau)} \nonumber \\
\mathcal{C}_2(\tau) &= -[\mathcal{C}_1(\tau)]^2-\tau\mathcal{C}_1(\tau) \nonumber \\
\mathcal{C}_3(\tau) &= -\tau \mathcal{C}_2(\tau)-2\mathcal{C}_1(\tau)\mathcal{C}_2(\tau)-\mathcal{C}_1(\tau) \nonumber \\
\mathcal{C}_4(\tau) &= -\tau \mathcal{C}_3(\tau)-2\mathcal{C}_2(\tau)-2\mathcal{C}_2^2(\tau)-2\mathcal{C}_1(\tau)\mathcal{C}_3(\tau) \nonumber \\
\mathcal{C}_5(\tau) &= -3\mathcal{C}_3(\tau)-\tau \mathcal{C}_4(\tau)-6\mathcal{C}_2(\tau)\mathcal{C}_3(\tau)-2\mathcal{C}_1(\tau)\mathcal{C}_4(\tau)
\label{C_form}
\end{align}

\begin{figure}[hbt!]
\centering
\includegraphics[scale=0.8]{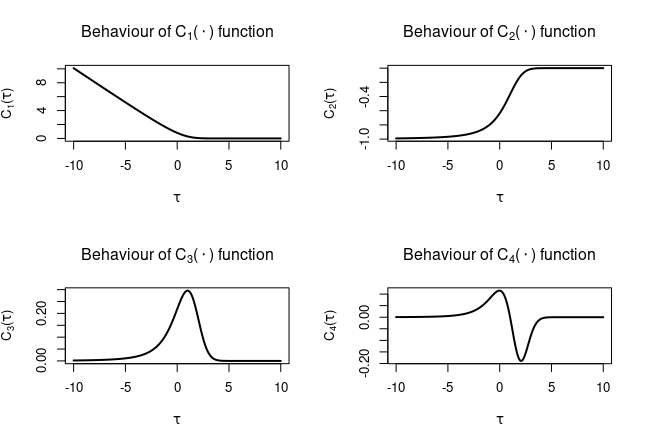}
\caption{Plotting C functions of an Extended Skew Normal distribution up to order four with respect to the $\tau$ parameter with range values $[-10,10]$.}
\label{ctau_fun}
\end{figure}

\noindent
Using~\eqref{C_form}, we can retrieve the constants associated to the moments of a Skew Normal random variable when $\tau = 0$, since  $\mathcal{C}_1(0)=\sqrt{\frac{2}{\pi}}$, $\mathcal{C}_2(0)=-\frac{2}{\pi}$, $\mathcal{C}_3(0)=\sqrt{\frac{2}{\pi}}\frac{(4-\pi)}{\pi}$ and $\mathcal{C}_4(0)=-\frac{24}{\pi^2}+\frac{8}{\pi}$. \\ The behavior of these $\mathcal{C}$ functions is shown in Figure \ref{ctau_fun} where we observe the following: 

\begin{description}
    \item[$\bullet$] $C_1(\tau)$ has a linear behavior for negative values and quickly decays to zero as $\tau$ approaches zero towards the positive range side
    \item[$\bullet$] $C_2(\tau)$ assumes values in the range $(-1, 0)$ and follows a logistic like shape
    \item[$\bullet$] $C_3(\tau)$ assumes values in the range $(0, 0.3)$ and resembles a probability density function
    \item[$\bullet$] $C_4(\tau)$ assumes values in the range $(-0.2, 0.1)$ and quickly decays to zero as $\tau < -1$ and $\tau > 4$
\end{description}

\noindent
In particular, the function $\mathcal{C}_3$ approximately satisfies all the required properties of a probability density function, the range is positive and the respective integral is close to one. \\ Numerical integration shows that the integral is 0.9991876 with absolute error less than 8.1e-05 for values of $\tau$ within the range [-35, 35]. This is helpful to simplify an implementation of the Extended Skew Normal distribution as there is no additional gain in considering large values of $\tau$. We return to this issue in Section \ref{sec:3}. \\ Both the additional hidden mean parameter and the $\mathcal{C}$ function patterns make the Extended Skew Normal distribution appealing for better modeling skewed posterior behaviors when properly encoded in the Simplified Laplace strategy.

\noindent
A closed expression for the parameterization $\delta$ is obtained from\eqref{esn_mom} as follows

\begin{equation}
\delta = \sign(\gamma_1) \sqrt{\frac{ |\gamma_1|^{2/3}}{[\mathcal{C}_3(\tau)]^{2/3}-\mathcal{C}_2(\tau)|\gamma_1|^{2/3}}}
\label{delta_tau}
\end{equation}

\noindent
Similar to existing solutions for the Skew Normal distribution, we may use the moments to construct a proper mapping for the Extended Skew Normal. We substitute equation~\eqref{delta_tau} into the kurtosis one in~\eqref{esn_mom} and achieve a solution for $\tau$. Then we see the following:

\begin{description}
\item[$\bullet$] the fourth equation with respect to $\tau$ does not have a closed form solution
\item[$\bullet$] the kurtosis is unbounded as its range is $[0, \infty)$ and this can lead to numerical issues or unreasonable outcomes
\end{description}

\noindent
As we do not control kurtosis results within a finite range, a mapping between parameters and moments of the Extended Skew Normal density moments is not feasible. \\ In Section \ref{sec:3} we show that it is way easier and more efficient to follow a similar scheme adopted for the Simplified Laplace strategy where we fit Skew Normal distributions by matching higher-order derivatives evaluated at the mode of the target distribution. The Extended Skew Normal distribution is no exception to this methodology since we only need an additional higher-order derivative to get solutions for the parameter $\tau$. This extended Skew Normal version can also be used to model skewness within the Latent Gaussian Model paradigm as it satisfies the Gaussian pattern discussed in Section \ref{sec:1}.

\subsection{Tail behaviour of the Skew Normal family}

Gaussian-like assumptions lead to accurate approximations of the posterior marginals of a Latent Gaussian Model in INLA. Apart from the Gaussian distribution, the Skew Normal family appears to be another natural choice for modeling these marginals. The Simplified Laplace strategy is built upon Skew Normal distributions, granting fast and accurate results when Gaussian assumptions are too limiting. Although the bulk of the distribution around the mode differs from a Gaussian due to its asymmetrical nature, we demonstrate here that both the Skew Normal distribution and its extended version satisfy the Gaussian-like posterior marginal representation discussed in Section \ref{sec:1}. Consider the log densities of a standard Skew Normal and Extended Skew Normal distribution 

\begin{align}
    \log f_{\SN}(x; \alpha) &= \log(2)+\log(\phi(x))+\log(\Phi(\alpha x)) \nonumber \\
    &= \log(2)+\log(\phi(x))+\log \Bigl(\frac{1}{2}+\frac{1}{2} \erf \Bigl (\frac{\alpha x}{\sqrt{2}} \Bigr)\Bigr ) 
    \nonumber \\
    \log f_{\ESN}(x; \alpha, \tau) &= -\log(\Phi(\tau))+\log(\phi(x))+\log(\Phi(\alpha x+\tau \sqrt{1+\alpha^2})) \nonumber \\
    &= -\log(\Phi(\tau))+\log(\phi(x))+\log \Bigl(\frac{1}{2}+\frac{1}{2} \erf \Bigl (\frac{\alpha x+\tau \sqrt{1+\alpha^2}}{\sqrt{2}} \Bigr)\Bigr )
    \label{eskew_err}
\end{align}
with $\erf(x)=\frac{2}{\sqrt{\pi}} \int_0^x \exp(-z^2) \, dz$ being the error function. We see that Gaussian distributions bound both densities since $\Phi(p) \le 1$. However, tail behavior is another important aspect of a distribution as it provides information of extreme observations. Considering Skew Normal family distributions as natural candidates for our deterministic marginal approximations, we need to ensure that even their tails follow a Gaussian behavior. This can be accomplished by computing series expansions of both log densities in~\eqref{eskew_err} for the limiting cases $x \rightarrow \pm \infty$. An asymptotic expansion of the log Gaussian density $\phi(x)$ is straightforward and consists of one squared term. Skew Normal family densities add more complexity because of the $\Phi(\cdot)$ function term. Asymptotic expansion results for both Skew Normal family tails are provided below, where $\nu = \tau \sqrt{1+\alpha^2}$. The results for the right tail are

\begin{align}
    \log f_{\SN}(x; \alpha) \vert_{x \rightarrow +\infty} &\approx -\frac{1}{2} x^2 +\exp \Bigl(-\frac{\alpha^2 x^2}{2} \Bigr) \Bigl (-\frac{1}{2}\frac{\sqrt{2}}{\alpha x \sqrt{\pi}}+\dots \Bigr ) \nonumber \\
    \log f_{\ESN}(x; \alpha, \tau) \vert_{x \rightarrow +\infty} &\approx -\frac{1}{2}x^2+\exp(-\frac{\alpha^2x^2}{2}-\nu \alpha x) \Bigl ( -\frac{1}{2} \frac{\sqrt{2}\exp(-\frac{1}{2}\nu^2)}{\sqrt{\pi} \alpha x}+\dots \Bigr ) 
    \label{right_tails}
\end{align}
while for the left tail we have

\begin{align}
    \log f_{\SN}(x; \alpha) \vert_{x \rightarrow -\infty} &\approx -\frac{1}{2}x^2(1+\alpha^2)+\log \Bigl( -\frac{1}{\alpha x \sqrt{2\pi}}\Bigr)+\dots \nonumber \\
    \log f_{\ESN}(x; \alpha, \tau) \vert_{x \rightarrow -\infty} &\approx -\frac{1}{2}(\alpha^2+1)x^2+\nu x +\log \Bigl ( -\frac{1}{2}\frac{\sqrt{2}\exp(-\frac{1}{2}\nu^2)}{\alpha x\sqrt{\pi}} \Bigr ) +\dots
    \label{left_tails}
\end{align}

\noindent
The expanded results in~\eqref{right_tails} show a sequence of higher-order terms that quickly approach zero as $x \rightarrow +\infty$. As expected, the right tail of both Skew Normal and Extended Skew Normal density gets more and more similar to the desired Gaussian one. Corresponding left tail results~\eqref{left_tails} for $x \rightarrow -\infty$ show a similar Gaussian pattern but with a slower decay. Here we recognize a log Gaussian density contribution with additional logarithmic terms coming from the expanded cumulative density $\Phi(\alpha x)$. \\ 
As discussed in Section \ref{sec:1}, these Skew Normal family densities appear to be a natural, reasonable choice to approximate Latent Gaussian posterior marginals as accurately as possible.

\section{The Simplified Laplace strategy using the Extended Skew Normal distribution}
\label{sec:3}

The Simplified Laplace strategy described in \cite{rue2009} is one of the most attractive choices to get posterior approximations of a Latent Gaussian Model structure as it essentially ensures fast computations with a negligible cost in accuracy for most of the cases. \\ This strategy applies a third-order Taylor expansion to the target posterior approximations. Then fits Skew Normal distributions by matching the expansion terms with the high order derivatives of its log-likelihood evaluated at the mode. \\ This section reviews the methodology behind this strategy, adding a new way to compute the required Skew Normal moments, which avoids further approximation and optimization steps for evaluating the mode (see also \cite{simon2020}). We propose to extend this whole approach by considering a fourth-order Taylor expansion and fit an Extended Skew Normal distribution which uses an additional hidden mean parameter $\tau$. \\ In this setting, we need to make sure this extended distribution ensures both robustness of the results and fast computational performances. 

\subsection{Third order Taylor expansion}

\begin{figure}[hbt!]
\centering
\includegraphics[scale=0.6]{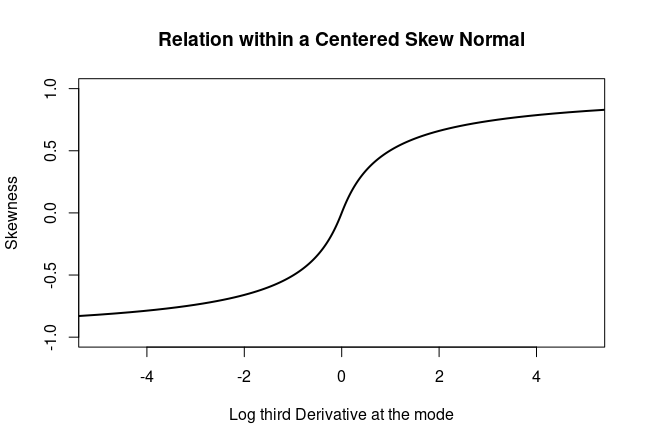}
\caption{The curve describes the exact relation of skewness and log third derivative evaluated at the exact mode of a standard Skew Normal random variable for many possible values of skewness in the range (-1,1). The modes are computed by numerical optimization for maximum accuracy purposes.}
\label{cen_deriv}
\end{figure}

The computational advantages of the Simplified Laplace strategy rely on accurate parametric density approximations instead of computing the more costly Laplace ones, which are based on a non-parametric representation of the posterior marginals. \\ The strategy consists of fitting a Skew Normal distribution to a third-order Taylor expanded density of the form

\begin{equation}
    \log (\pi(z)) = K-\frac{1}{2}z^2+\tilde{\mu}z+\frac{1}{3!}\tilde{\gamma}_1z^3+\dots
    \label{taylor_den}
\end{equation}
where $K$ is a constant, $(\tilde{\mu}, \tilde{\gamma}_1)$ are terms derived from the third order Taylor expansion of the Laplace Approximation. \\ The resulting density in~\eqref{taylor_den} is $N(\tilde{\mu}, 1)$ up to second order while the third term $\tilde{\gamma}_1$ provides information of the third order derivative evaluated at the mode. \\ Consider $R \sim \SN(\xi, \omega, \alpha)$ with unknown location $\xi$, scale $\omega$ and skewness parameter $\alpha$. Then we can define a system of three equations to compute the respective parameter triplet $(\tilde{\xi}, \tilde{\omega}, \tilde{\alpha})$ to approximate the expanded density in~\eqref{taylor_den}. By matching the first two non central moments and the third derivative of the Skew Normal at the mode $z^*$, the resulting system is 

\begin{align}
    &\E(R) = \tilde{\mu} \nonumber \\
    &\Var(R) = 1 \nonumber \\
    &\frac{\partial^3}{\partial r^3} \log \pi(r; \xi, \omega, \alpha) \Bigl \vert_{r=z^*} = \tilde{\gamma}_1
    \label{lin_sys}
\end{align}
However the mode $z^*$ is not analytically available. Following Appendix B of \cite{rue2009}, we can expand $\log \pi(r; \xi, \omega, \alpha)$ at its location point $r = \xi$ to compute an approximation to the mode as

\begin{equation}
    r^*=\Bigl ( \frac{\alpha}{\omega}\Bigr) \frac{\sqrt{2\pi}+2\xi \bigl ( \frac{\alpha}{\omega}\bigr)}{\pi+2\bigl ( \frac{\alpha}{\omega}\bigr)^2}
    \label{sn_mode}
\end{equation}
We evaluate the third derivative of the log Skew Normal density at the approximated mode~\eqref{sn_mode}. This expression is then expanded at $\frac{\alpha}{\omega}$ around $\alpha = 0$ to allow for an exact analytical result and fast computations. \\ We can avoid these steps and compute a more accurate modal configuration for a Skew Normal random variable using interpolation between skewness and third log derivative values. Figure \ref{cen_deriv} shows that the interpolation curve of the two quantities is smooth and can offer more precise results. \\ This interpolation avoids using an approximation for the mode. We use the interpolant to obtain the skewness and then compute the Skew Normal parameters directly from the corresponding mapping. \\ In most cases, we do not detect significant improvements but the new approach still makes the Simplified Laplace approximations slightly more accurate when non-negligible skewness is involved. Additionally, it simplifies the default INLA methodology avoiding computations for solving the system of equations. 

\noindent
The third equation in the system~\eqref{lin_sys} then becomes 

\begin{equation}
    \tilde{\gamma}_1 = \mathcal{C}_3(0) \Bigl (\frac{\alpha}{\omega} \Bigr )^3
    \label{third_sla}
\end{equation}
where the right side is exactly the resulting polynomial expansion of $\frac{\partial^3}{\partial r^3} \log \pi(r; \xi, \omega, \alpha) \Bigl \vert_{r=r^*}$ with $\mathcal{C}_3(\cdot)$ being the $\mathcal{C}$ function formulation derived from the Extended Skew Normal distribution. \\ Using equation~\eqref{third_sla}, we can directly solve the system~\eqref{lin_sys} since $\alpha$ is a function of the sole scale parameter $\omega$ with $\mathcal{C}_3(0)$ being a constant ($\approx$ 0.218).

\subsection{Fourth order Taylor expansion}

The Simplified Laplace methodology can be further extended by considering a fourth-order term in the expansion~\eqref{taylor_den}. In this framework, the Extended Skew Normal distribution described in Section \ref{sec:2} represents a natural match since it extends the Skew Normal distribution by introducing a fourth parameter in its analytical representation. \\ The corresponding log density of~\eqref{esn} can be written in a $\mathcal{C}$ function formulation as

\begin{equation}
\log f(t; \xi, \omega, \alpha, \tau) = \log \Bigl [ \frac{1}{\omega} \phi \Bigl (\frac{t-\xi}{\omega} \Bigr) \Bigr ]+ \mathcal{C}_0\Bigl( \tau \sqrt{1+\alpha^2}+\alpha \frac{t-\xi}{\omega} \Bigr) -\mathcal{C}_0(\tau)
\label{esn_den}
\end{equation}
If $\tau = 0$ the extended log density in~\eqref{esn_den} degenerates into a Skew Normal one. Moreover, the role of the hidden mean parameter  becomes irrelevant when $\alpha = 0$ as the density reverts back to a Gaussian distribution with mean $\xi$ and variance $\omega^2$. \\ According to \cite{seijas2017presence} and \cite{adelchiazzalini_2018_the},  $\tau$ affects both skewness and kurtosis of the distribution when $\alpha$ is not zero. It also determines the asymmetry of the density since it tends to 0 as $\tau \rightarrow \pm \infty$. \\ The log derivatives up to order four are the following 

\begin{align}
\frac{\partial}{\partial t} \log f(t; \xi, \omega, \alpha, \tau) &= -\frac{t-\xi}{\omega^2} + \mathcal{C}_1\Bigl ( \tau \sqrt{1+\alpha^2}+\frac{\alpha}{\omega} (t-\xi)\Bigr) \frac{\alpha}{\omega} \nonumber \\
\frac{\partial^2}{\partial t^2} \log f(t; \xi, \omega, \alpha, \tau) &= -\frac{1}{\omega^2} + \mathcal{C}_2\Bigl ( \tau \sqrt{1+\alpha^2}+\frac{\alpha}{\omega} (t-\xi)\Bigr) \Bigl (\frac{\alpha}{\omega} \Bigr)^2 \nonumber \\
\frac{\partial^3}{\partial t^3} \log f(t; \xi, \omega, \alpha, \tau) &= \mathcal{C}_3\Bigl ( \tau \sqrt{1+\alpha^2}+\frac{\alpha}{\omega} (t-\xi)\Bigr) \Bigl (\frac{\alpha}{\omega} \Bigr)^3 \nonumber \\
\frac{\partial^4}{\partial t^4} \log f(t; \xi, \omega, \alpha, \tau) &= \mathcal{C}_4\Bigl ( \tau \sqrt{1+\alpha^2}+\frac{\alpha}{\omega} (t-\xi)\Bigr) \Bigl (\frac{\alpha}{\omega} \Bigr)^4
\label{ext_der}
\end{align}

\noindent
We do not have an analytical solution for the mode due to the intractable structure of the first log derivative in~\eqref{ext_der}. \\ According to the Simplified Laplace methodology, we first expand the third log derivative at $t=\xi$ getting the new approximated mode

\begin{equation}
t^* = \Bigl ( \frac{\alpha}{\omega} \Bigl ) \frac{\mathcal{C}_1(\tau \sqrt{1+\alpha^2})-\mathcal{C}_2(\tau \sqrt{1+\alpha^2})\xi \Bigl ( \frac{\alpha}{\omega} \Bigl )}{1-\mathcal{C}_2(\tau \sqrt{1+\alpha^2}) \Bigl ( \frac{\alpha}{\omega} \Bigl )^2}
\label{esn_mode}
\end{equation}

\noindent
which reverts back to~\eqref{sn_mode} as $\tau = 0$. We chose not to use the interpolant function of Figure \ref{cen_deriv} for the Extended distribution since there are now two free parameters. Another existing numerical approximation for the mode is provided in \cite{adelchiazzalini_2018_the} by using the centralized moments of Skew Normal family densities. The final step sees the expansion of the third and fourth log derivatives of the Extended Skew Normal distribution at the mode~\eqref{esn_mode} with respect to $\frac{\alpha}{\omega}$ around $\alpha = 0$. Then we obtain two new polynomial approximations for these high order log derivatives 

\begin{align}
    \frac{\partial^3}{\partial t^3} \log f(t; \xi, \omega, \alpha, \tau) \Bigl \vert_{t=t^*} &\approx \mathcal{C}_3(\tau) \Bigl ( \frac{\alpha}{\omega} \Bigr )^3 \nonumber \\
    \frac{\partial^4}{\partial t^4} \log f(t; \xi, \omega, \alpha, \tau) \Bigl \vert_{t=t^*} &\approx \mathcal{C}_4(\tau) \Bigl ( \frac{\alpha}{\omega} \Bigr )^4
    \label{esn_poly}
\end{align}
that are available as functions of the scale parameter $\omega$, the skewness parameter $\alpha$ and the hidden mean parameter $\tau$. \\ The new system consists of four equations and is obtained by matching the first two moments of the Extended Skew Normal random variable and its higher-order expanded log derivatives in~\eqref{esn_poly} as follows

\begin{align}
\xi+\omega \delta \mathcal{C}_1(\tau) &= \tilde{\mu} \nonumber \\
\omega^2(1+\mathcal{C}_2(\tau) \delta^2) &= 1  \nonumber \\
\mathcal{C}_3(\tau) \Bigl ( \frac{\alpha}{\omega} \Bigr )^3 &= \tilde{\gamma}_1 \nonumber \\
\mathcal{C}_4(\tau) \Bigl ( \frac{\alpha}{\omega} \Bigr )^4 &= \tilde{\gamma}_2 
\label{esn_foureq}
\end{align}
with $(\tilde{\gamma}_1, \tilde{\gamma}_2)$ being the third and fourth log derivatives evaluated at the mode derived from the target approximated posterior in~\eqref{taylor_den}. \\ Lastly, we compute the solutions of the Extended Skew Normal parameters by solving the system~\eqref{esn_foureq}. No straightforward analytical solution is available and we must rely on interpolation to the parameter $\tau$.

\subsection{Interpolating the hidden mean solutions}

\begin{figure}[hbt!]
\centering
\includegraphics[width=1.0\textwidth]{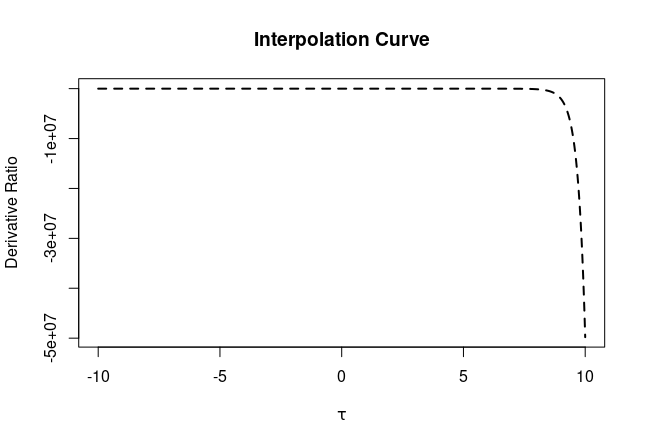}
\caption{Relationship between the truncation parameter $\tau$ and the $\mathcal{C}$ function derivative ratio $\frac{\mathcal{C}_4(\tau)}{[\mathcal{C}_3(\tau)]^{4/3}}$ obtained from \ref{tau_rel}. }
\label{mode_snesn}
\end{figure}

The Extended Skew Normal distribution can be used to fit a univariate target posterior approximation through a polynomial expansion up to order four. \\ The Simplified Laplace methodology describes how to get accurate results from a system of equations that involves matching moments and high order log derivatives of the new extended distribution. We observe that the last two equations in~\eqref{esn_foureq} lead to the following relation

\begin{equation}
    \frac{\tilde{\gamma}_2}{[\tilde{\gamma}_1]^{4/3}}=\frac{\mathcal{C}_4(\tau)}{[\mathcal{C}_3(\tau)]^{4/3}}
    \label{tau_rel}
\end{equation}
which is cumbersome to solve in terms of $\tau$ values. Nevertheless, Figure \ref{mode_snesn} shows there exists quite a smooth behaviour amongst the $\tau$ solutions for the $\mathcal{C}$ function ratio~\eqref{tau_rel}. \\ Instead of relying on costly non-linear solvers, we construct an interpolant function between $\tau$ and its derivative ratio within a reasonable range of solutions.

\noindent
The interpolant function ensures accurate and fast solutions for reasonable boundaries of $\tau$. We  can also notice that the derivative ratio is positively bounded from above as follows

\begin{equation}
- \infty < \frac{\mathcal{C}_4(\tau)}{[\mathcal{C}_3(\tau)]^{4/3}} < 2.4 \quad \text{with} \quad - \infty < \tau < \infty
\label{ratio_tau}
\end{equation}
which matters as an Extended Skew Normal distribution converges to a Gaussian distribution when $\tau \rightarrow \pm \infty$. \\ More precisely \cite{canale2011} shows that the limiting Gaussian cases are $N(\xi, \omega^2)$ for $\tau \rightarrow \infty$ and $N(-\alpha |\tau|, \frac{1}{\sqrt{1-\delta^2}})$ for $\tau \rightarrow - \infty$. \\ Since $\mathcal{C}_3(\tau)$ approximately resembles a probability density function with respect to the parameter $\tau$, we can consider a criterion to decide whether a resulting value of $\tau$ is reasonable or not according to the log derivative outcomes $(\tilde{\gamma}_1, \tilde{\gamma}_2)$. \\ As discussed in Section \ref{sec:2}, we establish that a $|\tau| > 10$ value is already far extreme and can lead to unlikely or unstable results. As the respective parameter probability space coverage given by $ \int_{-10}^{10} \mathcal{C}_3(\tau) \, d \tau \approx 0.99$ is high, this rule of thumb ensures to keep most of the solutions. \\ Additionally, a low value of $\tilde{\gamma}_1$ results in an unreasonable ratio outcome of \ref{ratio_tau} for the corresponding interpolant. When $\tilde{\gamma}_1$ approaches zero, the Extended Skew Normal density bends to a Gaussian one and the new approach gets unstable. \\ To account for these unreasonable scenarios we simply return to the original Simplified Laplace approach using Skew Normal distribution, if this happens. \\ Overall the interpolant for the hidden mean parameter $\tau$ works well and does not add computational costs. \\ \\ By exploiting interpolation to solve the ratio in~\eqref{tau_rel}, we can obtain solutions for the system derived from using an Extended Skew Normal distribution. \\ Assuming $\tilde{\gamma}_1$ is not zero, we write $a^*= \mathcal{C}_3(\tilde{\tau})$ where $\tilde{\tau}$ is the result obtained by the interpolant. We write the skewness parameter as $\tilde{\alpha}=\tilde{\omega} b^*$ with $b^*= (\frac{\gamma_3}{a^*})^(1/3)$ and get 

\begin{equation}
\tilde{\omega} = \sqrt{\frac{-d^*+\sqrt{(d^*)^2+4c^*\sigma^2}}{2c^*}}
\end{equation}

\noindent
where $c^*=(b^*)^2(1+\mathcal{C}_2(\tilde{\tau}))$ and $d^*=1-(b^*)^2$. If $\tilde{\tau}$ approaches $0$ then we revert to a Skew Normal system of equations. Here we know that the location $\tilde{\xi}$ is given by 

\begin{equation}
    \tilde{\xi} = \tilde{\mu}-\tilde{\omega}\tilde{\delta}\mathcal{C}_1(\tilde{\tau})
    \label{loc_esn}
\end{equation}
where $\tilde{\delta}=\frac{\tilde{\alpha}}{\sqrt{1+\tilde{\alpha}^2}}$. The last expression~\eqref{loc_esn} gives the final location solution for the Extended Skew Normal system. 

\section{Applications}
\label{sec:4}

Skew Normal family provides a class of parametric distributions that well approximate posterior marginals for Latent Gaussian Models. As discussed in Section \ref{sec:3}, we can use Skew Normal and Extended Skew Normal distributions to get deterministic approximations for these posteriors using INLA and its Simplified Laplace strategy. Since we are interested in comparing outcomes from different strategies from INLA and MCMC in more extreme cases, we set a series of simulations that trigger high marginal skewness. We expect to observe accuracy differences between the two parametric choices in this framework. We simulate data from Binomial and Poisson likelihoods with different sample sizes and one single covariate with Gaussian prior to keep things simple. We then proceed with a Bayesian inference analysis onto these GLM models by using the following strategies: the standard Simplified Laplace Approximation (SLA) with Skew Normal distributions, the extended Simplified Laplace Approximation (ESLA) using Extended Skew Normal distributions strategies, the full Laplace Approximation (LA) in INLA and the MCMC samples from JAGS.

\subsection{Comparing INLA and MCMC strategies}

\begin{figure}[bp!]
\centering
\includegraphics[scale=0.7]{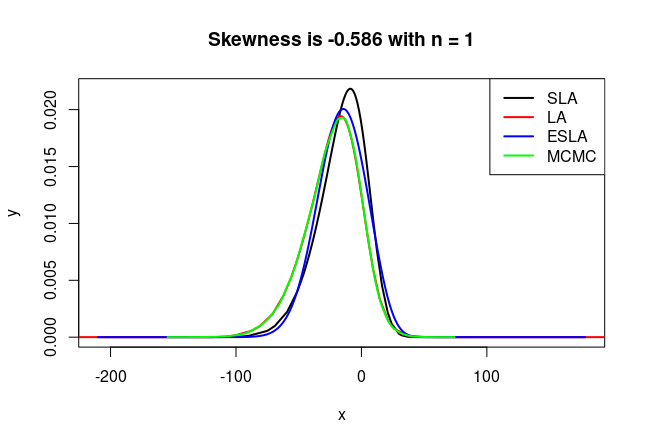}
\caption{Comparative results between SLA (black line), LA (red line), ESLA (blue line) and MCMC (green line) strategies with $n=1$ observations and a Bernoulli likelihood. Extreme negative skewness setting with minimum sample size. Since LA and MCMC strategies embody the posterior truth, we can observe that the SLA approach shows way less accuracy around the mode than its extended version denoted by ESLA. Tail behavior is similar for both SLA and ESLA and still appears to be slightly inaccurate in the left direction.}
\label{bin1}
\end{figure}

\begin{figure}[bp!]
\centering
\includegraphics[scale=0.7]{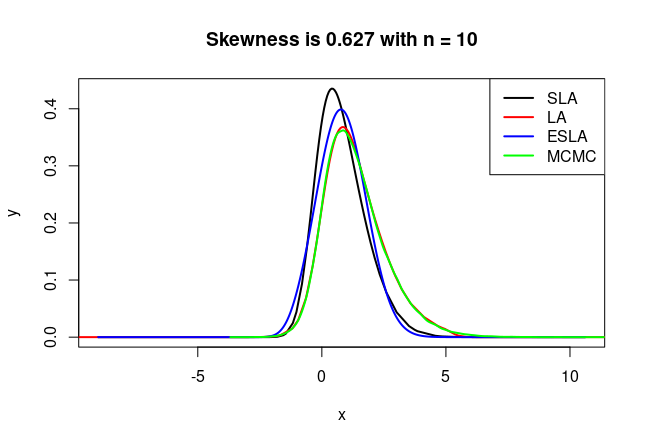}
\caption{Comparative results between SLA (black line), LA (red line), ESLA (blue line) and MCMC (green line) strategies with $n=10$ observations and a Bernoulli likelihood. Extreme positive skewness setting with small sample size. Since LA and MCMC strategies embody the posterior truth, we can see that the SLA approach shows way less accuracy around the mode than its extended version ESLA. Tail behavior is similar for both SLA and ESLA and still appears to be moderately inaccurate in the right direction.}
\label{bin10}
\end{figure}

\begin{figure}[bp!]
\centering
\includegraphics[scale=0.7]{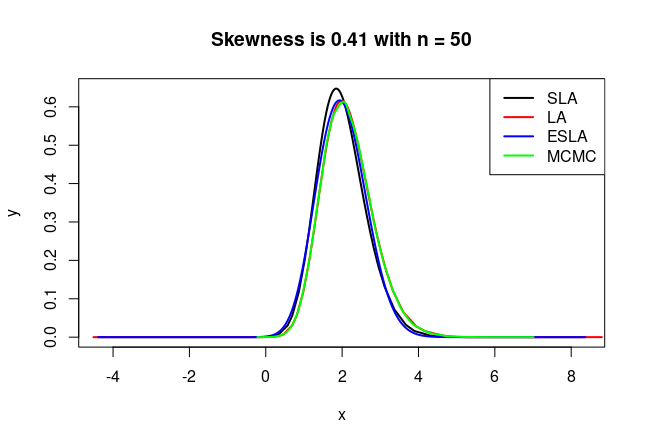}
\caption{Comparative results between SLA (black line), LA (red line), ESLA (blue line) and MCMC (green line) strategies with $n=50$ observations and a Bernoulli likelihood. Extreme positive skewness setting with moderate sample size. All employed strategies for this application show similar results except for the SLA methodology, which appears to be more inaccurate around the mode. Still, both SLA and ESLA suffer minor deviations in the right tail compared to LA and MCMC truth.}
\label{bin50}
\end{figure}

\begin{figure}[bp!]
\centering
\includegraphics[scale=0.7]{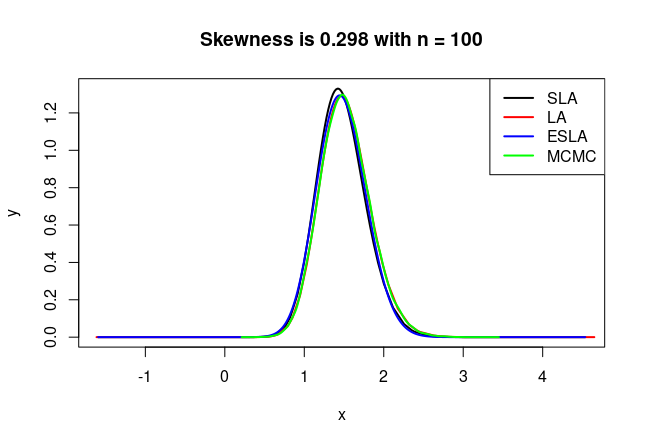}
\caption{Comparative results between SLA (black line), LA (red line), ESLA (blue line) and MCMC (green line) strategies with $n=100$ observations and a Bernoulli likelihood. High positive skewness setting with enough large sample size. All employed strategies for this application show similar results with minor deviations around the mode given by the SLA methodology. Large sample sizes tend to provide more stable expected results no matter the approximation strategy we use. Still, ESLA strategy is much closer to the true posterior results than SLA.}
\label{bin100}
\end{figure}

\begin{figure}[bp!]
\centering
\includegraphics[scale=0.7]{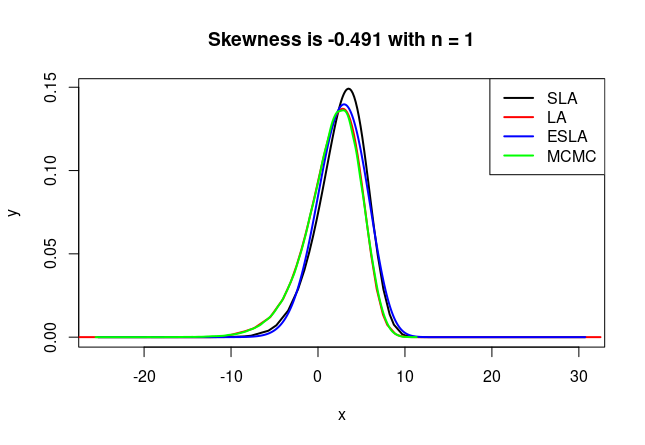}
\caption{Comparative results between SLA (black line), LA (red line), ESLA (blue line) and MCMC (green line) strategies with $n=1$ observations and a Poisson likelihood. Extreme negative skewness setting with minimum sample size. Since LA and MCMC strategies embody the posterior truth, we can observe that the SLA approach shows way less accuracy around the mode than its extended version denoted by ESLA. Unlike the Binomial case, tail behaviors for both SLA and ESLA closely match with no evident differences.}
\label{poi1}
\end{figure}

\begin{figure}[bp!]
\centering
\includegraphics[scale=0.7]{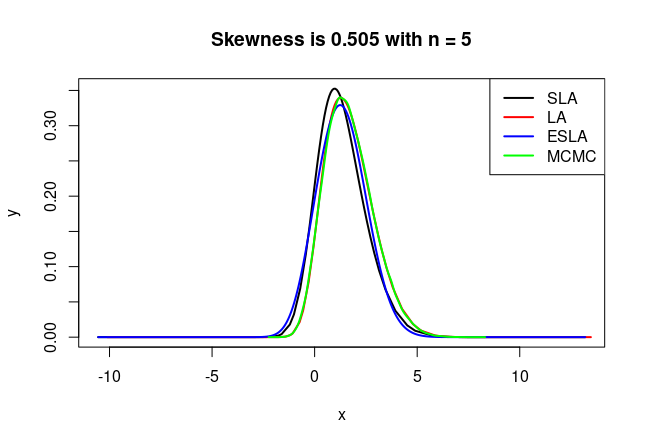}
\caption{Comparative results between SLA (black line), LA (red line), ESLA (blue line) and MCMC (green line) strategies with $n=5$ observations and a Poisson likelihood. Extreme positive skewness setting with small sample size. Since LA and MCMC strategies embody the posterior truth, we can see that the SLA approach shows way less accuracy around the mode than its extended version ESLA. Unlike the Binomial case, tail behaviors for both SLA and ESLA closely match with no evident differences.}
\label{poi5}
\end{figure}

\begin{figure}[bp!]
\centering
\includegraphics[scale=0.7]{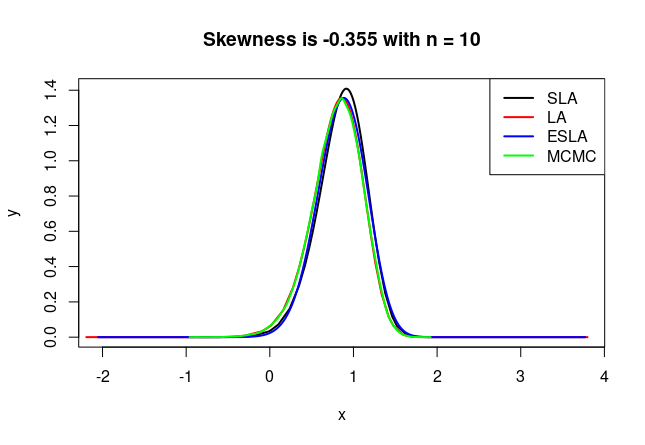}
\caption{Comparative results between SLA (black line), LA (red line), ESLA (blue line) and MCMC (green line) strategies with $n=10$ observations and a Poisson likelihood. High negative skewness setting with small sample size. All employed strategies for this application show similar results except for the SLA methodology, which appears to be more inaccurate around the mode. Still, both SLA and ESLA suffer minor deviations in the left tail compared to LA and MCMC truth.}
\label{poi10}
\end{figure}

\begin{figure}[bp!]
\centering
\includegraphics[scale=0.7]{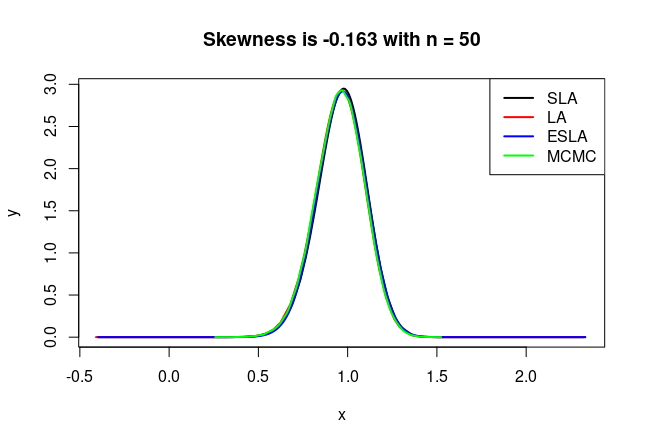}
\caption{Comparative results between SLA (black line), LA (red line), ESLA (blue line) and MCMC (green line) strategies with $n=50$ observations and a Bernoulli likelihood. Moderate negative skewness setting with enough large sample size. All employed strategies for this application closely converge to the same posterior result with no evident difference. Large sample sizes tend to provide more stable expected results no matter the approximation strategy we use.}
\label{poi50}
\end{figure}

The simulations for both Binomial and Poisson likelihoods are done with varying sample size dimensions $n$ from one observation up to 100. This setting results in non-negligible marginal skewness for the respective marginal posteriors. The posterior marginals tend to be less extreme when the sample size increases as they will converge to a Gaussian limit. \\ All resulting posterior marginals obtained from the different strategies are reported in the plots below. Comparison results from the Binomial model can be observed in Figure \ref{bin1}, \ref{bin10}, \ref{bin50} and \ref{bin100} while the Poisson ones are shown in Figure \ref{poi1}, \ref{poi5}, \ref{poi10} and \ref{poi50}. For low sample size $n$, we observe that the ESLA strategy provides more accurate results around the mode. The full Laplace (LA) and MCMC methods report the true results and do not differ in practice. ESLA posterior results appear closer to LA and MCMC than SLA strategy, where the mode is far off the expected location. \\ For larger sample size $n$, we tend to observe similar results for all strategies, with ESLA being slightly more accurate. A summary of the posterior modal configurations for different sample sizes is given on Tables \ref{bin_esla_mode} and \ref{poi_esla_mode}, while interquartile ranges (IQR) are reported in Table \ref{bin_esla_iqr} and \ref{poi_esla_iqr}. \\ These simulations underline that ESLA strategy is preferable in more extreme settings where the skewness is high. The extended methodology also preserves robustness as it is forced to revert back to a standard Simplified strategy in non-extreme cases.

\begin{table}[bp!]
\centering
\caption{Binomial simulations for increasing sample sizes up to $n=100$ and posterior mode evaluations using SLA, ESLA, LA and MCMC strategies. For low sample sizes, the modes derived from ESLA strategy are closer to the true ones from LA and MCMC approaches than the default SLA strategy. As the sample size $n$ increases, we notice a decreasing pattern for the positive skewness sequence (apart from $n=1$), with the mode values converging to the same result for all strategies. Overall, ESLA provides more coherent results to LA and MCMC, confidently representing the truth. }
\begin{tabular}{|l|l|l|l|l|l|}
\hline\noalign{\smallskip}
\textbf{n} & \textbf{Skew} & \textbf{Mode(SLA)} & \textbf{Mode(ESLA)}  & \textbf{Mode(LA)} & \textbf{Mode(MCMC)} \\
\noalign{\smallskip}\hline\noalign{\smallskip}
1 & -0.578 & -8.979 & -14.528 & -16.581 & -17.249  \\
2 & 0.644 & 0.346  & 0.783 & 1.084  & 0.995  \\
5 & 0.627 & 1.17 & 1.764 & 1.844 & 1.914 \\
10 & 0.495 & 1.207 & 1.39 & 1.459 & 1.363  \\
20 & 0.451 & 0.639 & 0.722 & 0.784  & 0.764  \\
50 & 0.306 & 0.908  & 0.934  & 0.964  & 0.94  \\
100 & 0.218 & 0.85  & 0.862  & 0.881  & 0.876  \\
\noalign{\smallskip}\hline
\end{tabular}
\label{bin_esla_mode}
\end{table}

\begin{table}[bp!]
\centering
\caption{Binomial simulations for increasing sample sizes up to $n=100$ and posterior interquartile range (IQR) evaluations using SLA, ESLA, LA and MCMC strategies. The IQRs from both SLA and ESLA strategies get closer and closer to the truth provided by LA and MCMC posterior results as soon as the sample size increases. Although the difference is less relevant than the one from the respective mode in Table \ref{bin_esla_mode}, ESLA grants more accurate results towards the truth than its simpler version SLA.}
\begin{tabular}{|l|l|l|l|l|l|}
\hline\noalign{\smallskip}
\textbf{n} & \textbf{Skew} &  \textbf{IQR(SLA)}  & \textbf{IQR(ESLA)}  & \textbf{IQR(LA)}  & \textbf{IQR(MCMC)}\\
\noalign{\smallskip}\hline\noalign{\smallskip}
1 & -0.578 & 25.865 & 26.909  & 28.73  & 28.949 \\
2 & 0.644  & 2.046  & 2.138  & 2.838  & 3.012 \\
5 & 0.627  & 2.316  & 2.4  & 2.751  & 2.80 \\
10 & 0.495  & 1.189  & 1.232  & 1.313  & 1.31 \\
20 & 0.451  & 0.755  & 0.78  & 0.813  & 0.816 \\
50 & 0.306  & 0.468  & 0.477  & 0.483  & 0.483 \\
100 & 0.218  & 0.365  & 0.37  & 0.372  & 0.372 \\
\noalign{\smallskip}\hline
\end{tabular}
\label{bin_esla_iqr}
\end{table}

\begin{table}[bp!]
\centering
\caption{Poisson simulations for increasing sample sizes up to $n=100$ and posterior mode evaluations using SLA, ESLA, LA and MCMC strategies. For low sample sizes, the modes derived from ESLA strategy are closer to the true ones from LA and MCMC approaches than the default SLA strategy. As the sample size $n$ increases, we notice a decreasing pattern for the negative skewness sequence (apart from $n=2$), with the mode values converging to the same result for all strategies. Overall, ESLA provides more coherent results to LA and MCMC, confidently representing the truth. }
\begin{tabular}{|l|l|l|l|l|l|}
\hline\noalign{\smallskip}
\textbf{n} & \textbf{Skew} & \textbf{Mode(SLA)} & \textbf{Mode(ESLA)}  & \textbf{Mode(LA)}  & \textbf{Mode(MCMC)} \\
\noalign{\smallskip}\hline\noalign{\smallskip}
1 & -0.446 & 0.972  & 0.905  & 0.87  & 0.886  \\
2 & 0.496 & -2.696  & -2.195  & -2.06  & -1.87  \\
5 & -0.322 & 1.882  & 1.85  & 1.822  & 1.814  \\
10 & -0.311 & 0.796  & 0.78  & 0.767  & 0.763  \\
20 & -0.223 & 0.992  & 0.983  & 0.973  & 0.969  \\
50 & -0.179 & 1.109  & 1.106  & 1.103  & 1.108  \\
100 & -0.113 & 1.033  & 1.032  & 1.03  & 1.026  \\
\noalign{\smallskip}\hline
\end{tabular}
\label{poi_esla_mode}
\end{table}

\begin{table}[bp!]
\centering
\caption{Poisson simulations for increasing sample sizes up to $n=100$ and posterior interquartile range (IQR) evaluations using SLA, ESLA, LA and MCMC strategies. The IQRs from both SLA and ESLA strategies get closer and closer to the truth provided by LA and MCMC posterior results as soon as the sample size increases. Although the difference is less relevant than the one from the respective mode in Table \ref{poi_esla_mode}, ESLA grants more accurate results towards the truth than its simpler version SLA.}
\begin{tabular}{|l|l|l|l|l|l|}
\hline\noalign{\smallskip}
\textbf{n} & \textbf{Skew}  & \textbf{IQR(SLA)}  & \textbf{IQR(ESLA)}  & \textbf{IQR(LA)}  & \textbf{IQR(MCMC)}\\
\noalign{\smallskip}\hline\noalign{\smallskip}
1 & -0.446  & 0.598  & 0.618  & 0.64  & 0.644 \\
2 & 0.496  & 3.588  & 3.717  & 3.936  & 3.92 \\
5 & -0.322  & 0.492  & 0.5  & 0.5  & 0.5 \\
10 & -0.311  & 0.251  & 0.256  & 0.256  & 0.257 \\
20 & -0.223  & 0.224  & 0.227  & 0.227  & 0.227 \\
50 & -0.179  & 0.092  & 0.093  & 0.093  & 0.093 \\
100 & -0.113  & 0.082  & 0.083  & 0.083  & 0.083 \\
\noalign{\smallskip}\hline
\end{tabular}
\label{poi_esla_iqr}
\end{table}

\section{Discussion}
\label{sec:5}

Latent Gaussian Models provide an appealing hierarchical model structure for Bayesian inference as the a priori Gaussian assumption binds the posterior marginals. We discussed that densities with Gaussian tails can be a natural choice for approximating these marginals. Under these assumptions, the INLA methodology works well for this class of models by constructing fast and accurate deterministic approximations. Among the different available approximation options, the Simplified Laplace strategy is indeed one of the most advantageous for its speed and accuracy trade-off. This strategy relies on Skew Normal approximations of a third-order Taylor series expansion of the Laplace approximations, the latter of which are known to be highly accurate but computationally demanding. Skew Normal densities satisfy the Gaussian tail argument for modeling the latent posterior marginals while allowing non-negligible skewness.\\
However, this parametric assumption can pose a limit in more extreme cases, and we questioned if a more appropriate solution can be formulated. We chose another natural parametric distribution that still belongs to the Skew Normal family and ensured the Gaussian bounds are preserved: the Extended Skew Normal distribution. As reported in \cite{adelchiazzalini_2018_the}, this distribution is one of the Skew Normal extensions that has an additional parameter that affects all the moments, but in particular, the skewness for our purposes. Like the Simplified Laplace strategy, we formulated a system of equations by matching higher order derivatives of the Extended Skew Normal distribution evaluated at the mode, with the respective ones obtained from the expanded Laplace approximations. By interpolating some of the fourth parameter $\tau$ solutions, we efficiently calculate all four parameters necessary to fit an Extended Skew Normal approximation to the expansion. This alternative parametric approximation extends the capabilities of the Simplified Laplace strategy offering more accurate skew marginals, especially in more extreme settings. This work contributes an additional accurate and computational efficient approximation within the INLA framework, based on the Extended Skew Normal distribution and innovative solutions to calculate the necessary parameters. We believe that this contribution enables more accurate but still efficient Bayesian inference of complex models in the statistical community as well as the scientific community at large.

\begin{appendices}

\section{A special case: t-student as a normal mixture}
\label{AppA}

The Gaussian distribution provides bounds for the posterior marginals of a generic Bayesian inference up to a constant (see Section \ref{sec:1}). This is even more clear when the observed data $y_1, \dots, y_n$ are Gaussian distributed since the constants follow the same pattern. There are other cases that may show a non normal behaviour but they can still be cast into a Latent Gaussian paradigm. As an example, the t-student distribution is a statistical representation that allows for normal mixture structure but one can also consider logistic and Laplace distributions as well. As reported in Chapter 4 in \cite{held2005}, t-student assumptions can be encoded through a \emph{scale mixture of normals} by having $\boldsymbol{x} \vert \boldsymbol{\lambda} \sim N(\boldsymbol{0}, \boldsymbol{\lambda}^{-1}\boldsymbol{Q}^{-1})$  with $\boldsymbol{x}$ being a latent field component and $\boldsymbol{\lambda}$ a diagonal matrix of auxiliary variables. Introducing such auxiliary variables into the hierarchical representation of the latent model eases the overall structure when non normal assumptions are involved. Combining auxiliary variables and t-student information lead to the so called \emph{hierarchical t-formulation}. In particular, we underline the case where we assume the latent field $\boldsymbol{x}$ to be t-student distributed with normal data $\boldsymbol{y}$. As we employ auxiliary variables to get a normal scale structure, its hierarchical t-representation would be as follows

\begin{align}
\boldsymbol{y} \vert \boldsymbol{x} &\sim \prod_{i=1}^n \pi(y_i \vert x_i) \nonumber \\
\boldsymbol{x} &\sim \text{N}(\boldsymbol{0}, \boldsymbol{\lambda}^{-1} \boldsymbol{Q}^{-1}) \nonumber \\
\diag \{ \boldsymbol{\lambda} \} &\sim \prod_{i=1}^n \text{G} \Bigl (\frac{a_i}{2}, \frac{a_i}{2} \Bigr )
\end{align}
where $\pi(\boldsymbol{y} \vert \boldsymbol{x})$ is Gaussian while the mixing parameters $\lambda_1, \dots, \lambda_n$ are Gamma distributed. We assume our likelihood to be bounded by constants $\tilde{K}_1,\dots,\tilde{K}_n$, and end up with the following joint posterior relation

\begin{equation}
\pi(\boldsymbol{x}, \boldsymbol{\lambda} \vert \boldsymbol{y}) \propto \pi(\boldsymbol{y} \vert \boldsymbol{x}) \pi(\boldsymbol{x} \vert \boldsymbol{\lambda}) \pi(\boldsymbol{\lambda}) \le \pi(\boldsymbol{x} \vert \boldsymbol{\lambda}) \pi(\boldsymbol{\lambda}) \tilde{K}
\label{scale_t}
\end{equation}
where $\tilde{K} = \prod_i \tilde{K}_i$ is an overall constant. Then the full conditional is bounded as 

\begin{equation}
\pi(\boldsymbol{x} \vert \boldsymbol{\lambda}, \boldsymbol{y}) \le \tilde{K} \pi(\boldsymbol{x} \vert \boldsymbol{\lambda})
\label{t_full}
\end{equation}

\noindent
Similarly to the derivation in Section \ref{sec:1}, we obtain a bound for the corresponding marginals

\begin{equation}
\pi(x_i \vert \lambda_i, \boldsymbol{y}) \le \tilde{K}_i \exp \Bigl ( -\frac{ \lambda_i x_i^2}{2}(\boldsymbol{Q}^{-1})_{ii}^{-1}\Bigr ),
\label{t_part}
\end{equation}

\noindent
that are again bounded by a Gaussian distribution. The inequalities~\eqref{t_full} and~\eqref{t_part} show that we have control on all possible full conditional densities of the model as they are bounded by Gaussian densities. The same does not apply to the marginals $\pi(x_i \vert \boldsymbol{y})$ since they would still be bounded by t-student distributions. Non-normal assumptions on latent field or likelihood add complexity in approximating posterior marginals from these hierarchical structure.\\ The mixture representation of marginal posterior densities~\eqref{inla_mixt} entirely depends on full conditionals as we integrate out all the hyperparameters. Both the parametric and non parametric strategies of the methodology will still provide accurate results when the latent field is not normal.

\end{appendices}

\bibliographystyle{apalike}
\bibliography{References.bib}

\end{document}